# Seasonal thermal stress analysis of defective mass concrete sidewalls based on the average forming temperature method

Ziyan Zhao[a], Ting Peng[a], Peng Wu[a], Chaojun Hu[b], Qilin Yi[b], Chuangrui Huang[a], Junjie Niu[a], Xiaoxue Xu[a], Tao Li[a], Yuan Li[a, *]

[a] *Key Laboratory for Special Area Highway Engineering of Ministry of Education, Chang'an University, Xi'an 710064, China*

[b] *Sichuan Chuanjiao Road & Bridge Co., Ltd, De'yang 618000, China*

*E-mail: liyuan_mm@chd.edu.cn(Y.Li)

ABSTRACT

Thermal cracking in urban underground sidewalls is frequently observed when structures are cast in summer and enter service in winter, as seasonal temperature gradients act under structural restraint. To quantify the local stress field associated with pre-existing cracks, an orthogonal finite-element simulation matrix of 16 combinations is constructed. Distributions of maximum principal stress ($\sigma_1$) at the surface crack tip and along the upper half of the crack bottom are evaluated using steady-state thermal loading and a linear-elastic constitutive model. Across all cases, pronounced tensile stress concentration occurs at both locations: the maximum $\sigma_1$ ranges from 19.2 to 34.1 MPa at the crack surface end and from 17.2 to 29.4 MPa at the crack bottom. These concentrated values are consistently higher than the stress level at the same locations in an otherwise identical uncracked wall, clarifying how seasonal temperature gradients under restraint amplify local stresses around existing defects. The quantitative ranges reported here provide a basis for risk screening and for formulating practical mitigation measures (e.g., joint spacing and insulation strategies) in the design and operation of urban underground enclosure walls. In addition, three-dimensional simulations of randomly distributed internal voids show that adopting average forming temperature increases the peak tensile stress on void surfaces from 3.42 to 4.40 MPa at 10 °C and from 5.98 to 6.96 MPa at −5 °C, further highlighting the risk amplification effect of AFT under cold service conditions.

## 1. Introduction

The rapid urbanization in China has led to accelerated infrastructure development, significantly increasing the scale and complexity of engineering projects such as railways, highways, bridges, and subways. In this context, mass concrete structures are widely used in modern infrastructure as essential construction materials. Typical applications include concrete bridges, large-scale dams, basement slabs of high-rise buildings, and foundation platforms for heavy machinery. These implementations demonstrate mass concrete's advantages in delivering structural strength and stability, while highlighting its irreplaceable role in large-scale engineering projects[1-3].

There is no globally unified standard for mass concrete, with specifications varying among national codes. China's "Code for Construction of Mass Concrete" defines it as concrete elements with minimum dimensions over 1 meter, or those where hydration-induced temperature variations and shrinkage may cause detrimental cracking.

Current research on thermal cracking in mass concrete, both domestic and international, primarily focuses on the early hydration phase. The substantial dimensions of mass concrete structures lead to significant heat generation during cement hydration after placement, causing rapid temperature rise within the structural core. Meanwhile, faster surface cooling through air convection establishes pronounced thermal gradients between the core and surface regions, generating substantial thermal stresses[4]. When these stresses exceed the early-age concrete strength, cracking becomes inevitable, compromising structural integrity and durability[5, 6].

In contrast, long-term thermal stresses induced by environmental temperature variations during operational phases are frequently underestimated or neglected. Seasonal temperature fluctuations after concrete placement cause cyclic thermal variations within structural elements. Constrained thermal expansion/contraction deforming generate significant stress concentrations when free movement is restricted. Particularly under "summer casting-winter service" conditions, cooling from elevated initial placement temperatures combined with external restraints may induce critical thermal stresses that threaten long-term structural safety[7-10].

Current simplified approaches for seasonal thermal stress analysis often employ monthly average temperatures during casting/service periods as differential inputs, crudely assuming ambient casting temperature as the zero-stress reference or equating initial placement temperature with stress-free state. These oversimplifications fail to capture actual stress evolution, potentially yielding underestimated thermal stress predictions[7, 10-12].

The casting process of mass concrete inherently produces macro-defects such as internal voids and surface cracks due to complex interactions between material properties, construction techniques, and environmental conditions. Such defects substantially compromise structural stability and durability, posing latent risks to concrete integrity. This study therefore investigates thermal stress evolution in pre-existing concrete defects under seasonal temperature variations based on "Average Forming Temperature (AFT)[13] " theory, establishing critical foundations for developing preventive and remedial measures to ensure long-term structural stability—a research endeavor with profound theoretical and practical significance.

## 2. Methodology

This study investigates thermal stress in defective mass concrete based on AFT, with a focus on the effects of internal void defects and surface cracks [14].

*2.1. Research Methodology for Internal Defects*

*2.1.1 Stochastic Internal Defect Modeling*

Suppose there is a large-volume concrete block with dimensions of 1 m×1 m×1 m. To simulate the defect conditions during the pouring process of the concrete block, 100 spherical holes with a diameter of 1cm are randomly generated inside the concrete block structure. The corresponding defect model is then constructed. The following shows the code used to generate this defect model:

```
Inputs: N; Lx, Ly, Lz; t; rmin, rmax
Outputs: H = {(xk, yk, zk, rk)} and geometry = DIFFERENCE(BLOCK, union(SPHERE_k))

1:  H <- {}
2:  Sampling box: X in [t, Lx - t], Y in [t, Ly - t], Z in [t, Lz - t]
3:  while |H| < N do
4:      repeat
5:          Sample (hx, hy, hz) uniformly from the box
6:          Sample hr uniformly from [rmin, rmax]
7:          overlap <- false
8:          for each (xi, yi, zi, ri) in H do
9:              if sqrt((hx-xi)^2 + (hy-yi)^2 + (hz-zi)^2) < (hr + ri) then
10:                 overlap <- true; break
11:             end if
12:         end for
13:     until overlap = false
14:     Append (hx, hy, hz, hr) to H
15: end while
16: Create BLOCK(Lx,Ly,Lz) at origin
17: For each (xk, yk, zk, rk) in H, create SPHERE_k at (xk, yk, zk) with radius rk
18: S <- union of all SPHERE_k
19: Return geometry = DIFFERENCE(BLOCK, S), H
```

Run the code in the above table, and the generated model results are as Fig 1:

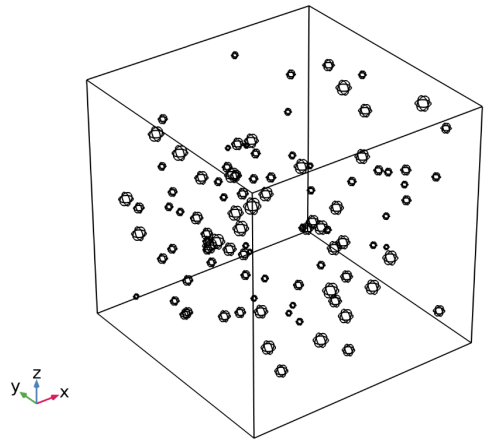

Fig 1. Internal Hole Defect Model

*2.1.2 Stress field solution*

In this simulation, the concrete mix ratio and thermal parameters are based on the values for C40 concrete as shown in Table 1. Since this is a thermo-mechanical coupled simulation, it is necessary to set the coefficient of thermal expansion, modulus of elasticity, and Poisson's ratio[15]. The coefficient of thermal expansion of concrete, $\alpha_T$ is $10^{-5}\,°C^{-1}$, the modulus of elasticity, $E$ is 30GPa, and the Poisson's ratio, $v$ is 0.2. The convective heat transfer coefficient on the outer surface of the sidewall model, in contact with the external environment, is set to $\beta_1$ is $22\,W/(m^2·°C)$. The convective heat transfer coefficient on the inner surface is typically lower, and the coefficient, $\beta_2$ is $15 W/(m^2 \cdot °C)$, to reflect the different heat exchange characteristics of the inner and outer surfaces. The initial temperature when entering the mold, $T_0$ is set to 30°C.

The external environmental temperature is represented by a cosine function based on the daily temperature variation of the local area: $T_a = 30 + 8\cos(\pi/12 t + \pi/4)$ °C to accurately simulate the dynamic changes in the actual environmental temperature. Sliding constraints are applied on both sides of the model to restrict longitudinal deformation, while allowing free lateral deformation, in order to simulate the boundary conditions of the sidewall in actual engineering[7, 10].

Suppose the concrete is poured during the sweltering summer days. Calculate the AFT field and designate it as the strain-free temperature field. Subsequently, simulate and obtain the distribution maps of the maximum principal stress (σ₁) under winter ambient temperatures of 10 °C, 5 °C, 0 °C, and -5 °C.

Meanwhile, for the sake of comparative analysis, an additional scenario is added to solve the stress field without taking the AFT.

Table 1
Parameter Table of Concret

| strength | $m_0$ | $m_1$ | $m_2$ | $Q_\infty$ | k | $Q_\infty^*$ | c | $\rho$ | $Q_{max}$ |
|---|---|---|---|---|---|---|---|---|---|
| C40 | 295 | 60 | 40 | 377 | 0.955 | 360 | 960 | 2400 | 61.7 |

The density of the C40 strength concrete specimen is $\rho = 2400 kg/m^3$, and the specific heat capacity is $c = 960 J/(kg \cdot °C)$. The cement used in the concrete mix has a final hydration heat $Q_\infty = 377 kJ/kg$, and the amount of cement is $m_0 = 295 kg$; the supplementary materials include fly ash $m_1 = 60 kg$ and slag $m_2 = 40 kg$. Based on equations (1) and (2), the reduced hydration heat release $Q_\infty^* = 360 kJ/kg$ is obtained. Then, according to equation (3), the maximum adiabatic temperature rise $\theta_{max} = 61.72°C$ can be calculated.

The heat release per unit volume of concrete is expressed as follows:

$$Q_\infty^* = k \cdot Q_\infty \tag{1}$$

In the formula:

$Q_\infty^*$—heat release per unit mass of concrete, $kJ/kg$;

$Q_\infty$—final heat release per unit mass of cement, $kJ/kg$;

k—reduction coefficient。

The reduction coefficients are different for different dosages of supplementary materials. In the concrete mix, when both fly ash and slag powder are used, the reduction coefficient can be calculated using the following formula:

$$k = k_1 + k_2 - 1 \tag{2}$$

In the formula:

$k_1$—reduction coefficient corresponding to the dosage of fly ash, the values are as shown in Table 2;

$k_2$—reduction coefficient corresponding to the dosage of slag, the values are as shown in Table 2.

$$\theta_{max} = WQ_\infty^*/c\rho \tag{3}$$

In the formula:

$\theta_{max}$ —The maximum adiabatic temperature rise of concrete shall be determined by adopting the values specified in Table 2, °C;

$Q_\infty^*$ —Maximum heat release per unit mass of concrete, $kJ/kg$;

$W$ —Dosage of cementitious materials per cubic meter of concrete, kg/m³;

$c$ —Specific Heat Capacity of Concrete; $J/(kg \cdot ℃)$;

$\rho$ —Density of Concrete, $kg/m^3$。

Table 2
Reduction coefficients of supplementary materials for different dosages

| dosage (%) | 0 | 10 | 20 | 30 | 40 |
|---|---|---|---|---|---|
| $k_1$ | 1 | 0.96 | 0.95 | 0.93 | 0.82 |
| $k_2$ | 1 | 1 | 0.93 | 0.92 | 0.84 |

*2.2. Surface Defects*

2.2.1 Development of Surface Crack Models
Tunnel lining structures are continuous along the longitudinal direction, and full three-dimensional finite element modeling of the entire tunnel is computationally expensive. Therefore, this study focuses on a representative tunnel sidewall segment with a longitudinal length of 10 m. Thus, this study focuses on the tunnel sidewall (longitudinal dimension: 10 m). Per Code, the maximum allowable crack width for concrete structures is 0.2 mm; a vertical 0.2 mm-wide crack is created at the outer surface center[16], as shown in the Fig 2.

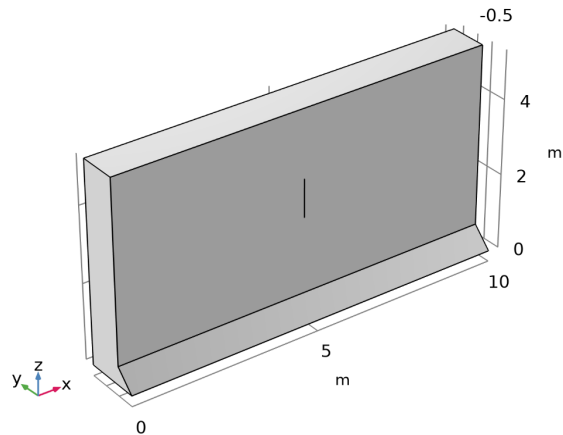

Fig 2. Concrete model.

2.2.2 Orthogonal Experimental Design Scheme
Orthogonal experimental design is a method used to study multiple factors at various levels. By systematically arranging multi-factor experiments, it utilizes orthogonal tables to select representative experimental combinations. This method can significantly reduce the number of experiments, save time and cost, and comprehensively and efficiently analyze the influence patterns of various factors[17, 18]. Orthogonal design experiments enable the identification of the key factors affecting existing cracks and clarification of the primary–secondary relationships and interactions among these factor affecting existing cracks, clarify the primary - secondary relationships and interactions among these factors, which provides a basis for the disease assessment of concrete structures, the improvement of their durability, and the formulation of targeted repair and reinforcement strategies. The casting temperature serves as the initial condition for temperature field Forming during concrete curing[19, 20].

The placement temperature is the initial condition for the development of the temperature field during the concrete forming (early-age) stage. It determines the baseline temperature at the onset of hydration heat release, directly influencing subsequent temperature accumulation processes. As a controllable construction parameter, it can be effectively regulated through adjustments to mixing temperature, reduction of material storage temperature, or control of transportation time. Cracks, acting as geometric discontinuities within concrete, directly affect localized heat conduction.

Their length or depth may alter local temperature gradients, thereby modifying heat accumulation patterns and average curing temperature distributions during the casting phase.

Therefore, three primary influencing factors were selected: casting temperature of concrete sidewalls (denoted as Factor A), existing crack length (Factor B), and existing crack depth (Factor C).

Each factor was assigned four levels (designated by numbers 1–4), as presented in Table 3.

Table 3
Factors and factor levels of the orthogonal experiment

| Factor | level | | | |
|---|---|---|---|---|
| | 1 | 2 | 3 | 4 |
| A(°C) | 25 | 30 | 35 | 40 |
| B(cm) | 10 | 30 | 50 | 70 |
| C(mm) | 2 | 3 | 4 | 5 |

The orthogonal array $L_{16}(3^4)$ was adopted for this three-factor, four-level experimental design.

For the three-factor, four-level experimental design, the orthogonal array $L_{16}(3^4)$ was selected to generate 16 experimental combinations. Under the premise of maintaining constant other conditions, for each combination, the average curing temperature field of concrete was calculated and set as the reference temperature. Based on this, seasonal temperature variation simulations were conducted under winter ambient temperature conditions of 5°C to obtain the stress distribution in the concrete structure for each combination.

## 3. Result

### 3.1. Results of the Solution of the Stress Field of the Random Model of Internal Defects

By solving the results in two cases, it is found that the stress concentration phenomenon will occur on the surfaces of all internal holes, and the stress concentration areas are in the shape of rings surrounding the holes. In order to observe the characteristics of the stress distribution more clearly, a certain hole model is extracted separately, and the distribution of the stress results at different temperatures is shown in Fig 3.-Fig 6

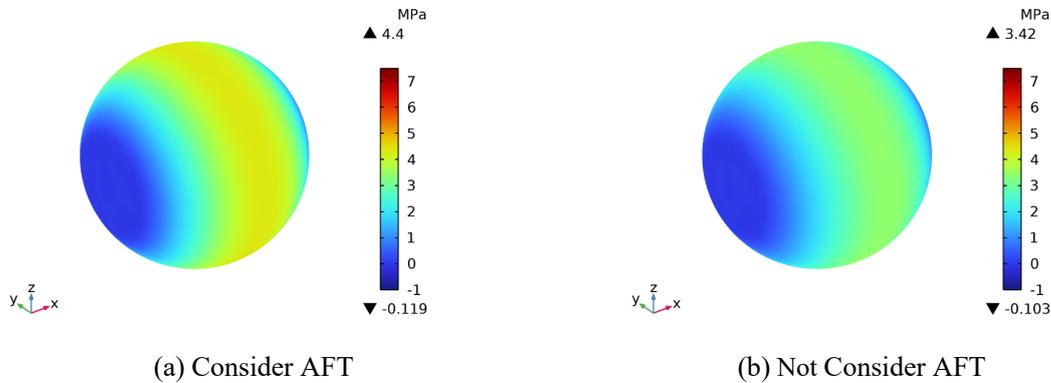

(a) Consider AFT          (b) Not Consider AFT

Fig 3. Distribution of the maximum principal stress ($\sigma_1$) on the hole surface at an ambient temperature of 10 °C.

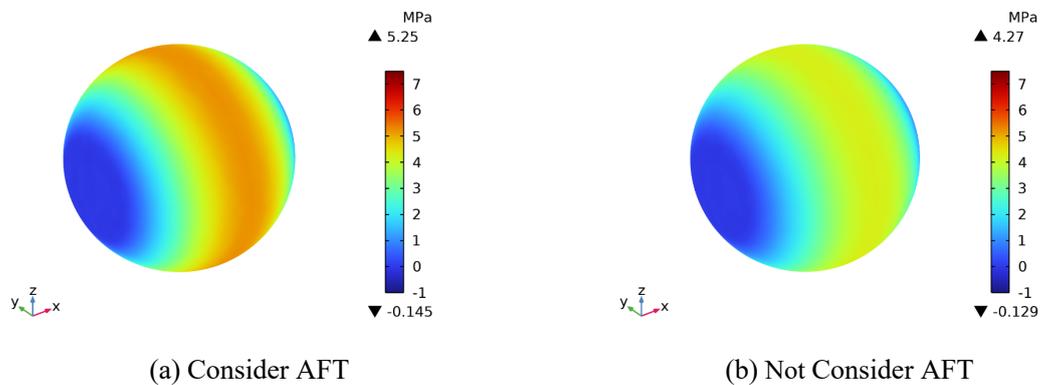

(a) Consider AFT          (b) Not Consider AFT

Fig 4. Distribution of the maximum principal stress ($\sigma_1$) on the hole surface at an ambient temperature of 5 °C.

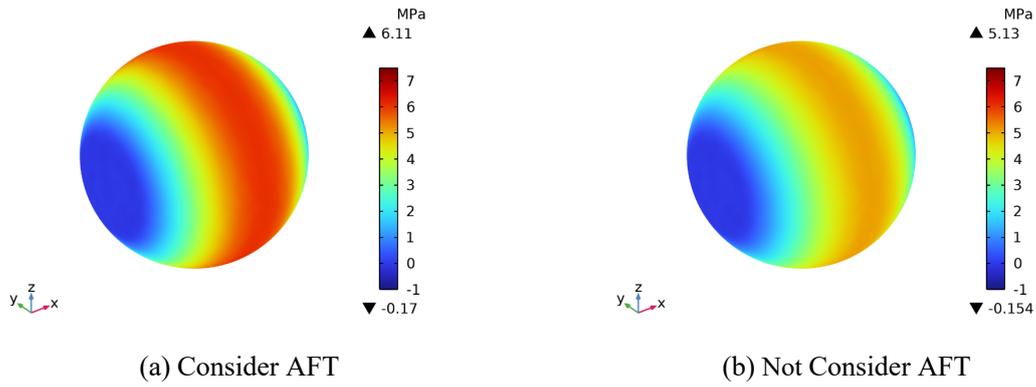

(a) Consider AFT　　　　　　　　　　　　　　(b) Not Consider AFT

Fig 5. Distribution of the maximum principal stress ($\sigma_1$) on the hole surface at an ambient temperature of 0 °C.

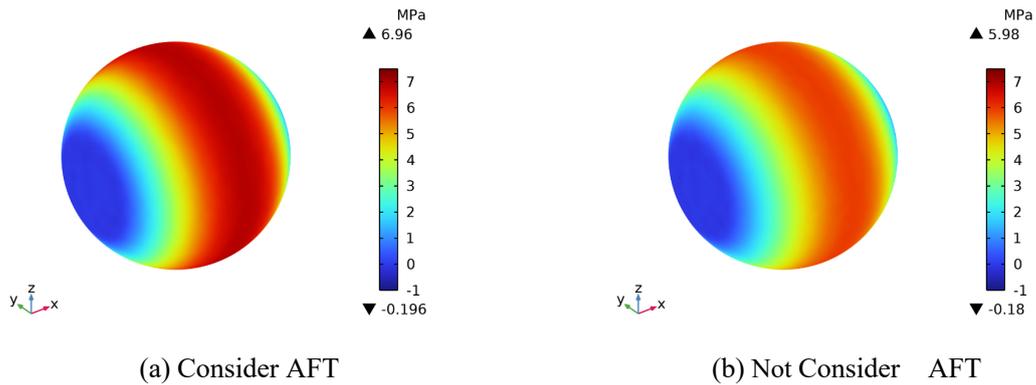

(a) Consider AFT　　　　　　　　　　　　　　(b) Not Consider　AFT

Fig 6. Distribution of the maximum principal stress ($\sigma_1$) on the hole surface at an ambient temperature of -5 °C.

As shown in the figures, under the same low-temperature environment, the first principal stress on void surfaces is significantly higher when AFT is considered (vs. non-considered cases). In both scenarios, stress increases with decreasing temperature. This indicates that under external low temperatures and boundary constraints, concrete is prone to cracking along internal void surfaces, with cracks propagating further—negatively impacting structural stability and threatening safety.

*3.2. Results of the Solution of the Stress Field of the Random Model of Surface Defects*

Simulation results indicate obvious stress concentration at the crack tip and bottom across all test parameter combinations, with local stress significantly higher than in surrounding areas. To illustrate this distribution, Fig 7 to Fig 8 show the first principal stress distribution at the crack tip and upper half bottom for 16 test combinations.

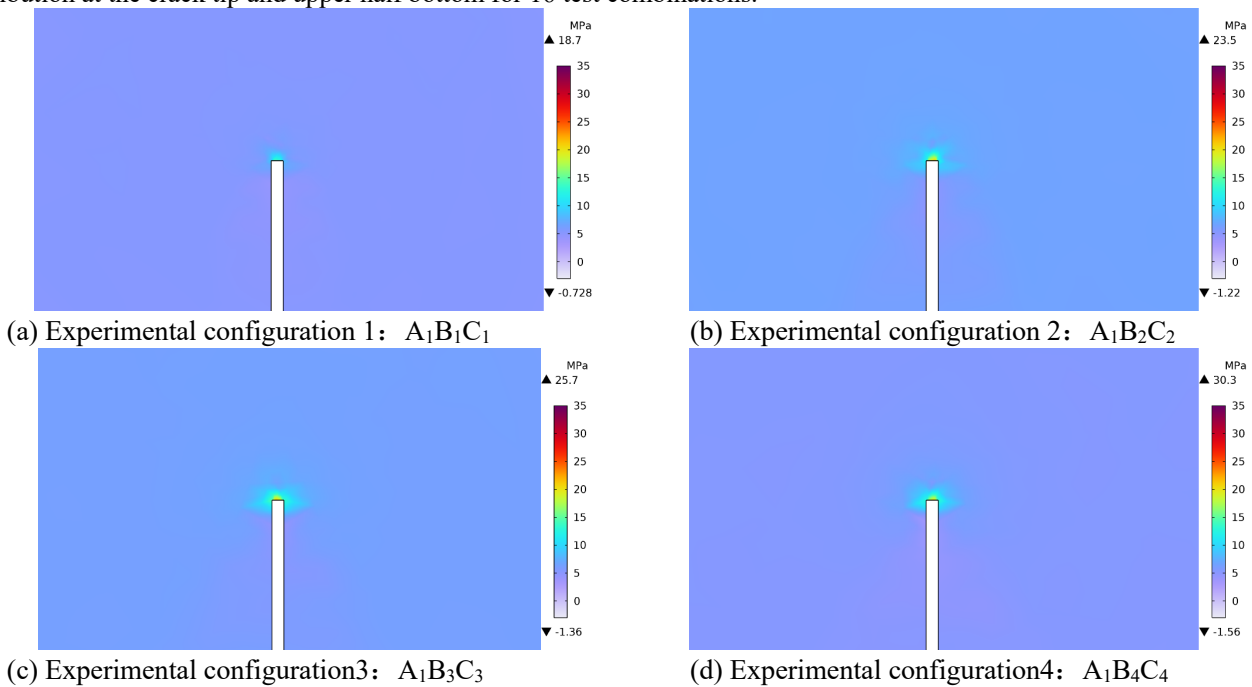

(a) Experimental configuration 1：$A_1B_1C_1$　　　　　(b) Experimental configuration 2：$A_1B_2C_2$

(c) Experimental configuration3：$A_1B_3C_3$　　　　　(d) Experimental configuration4：$A_1B_4C_4$

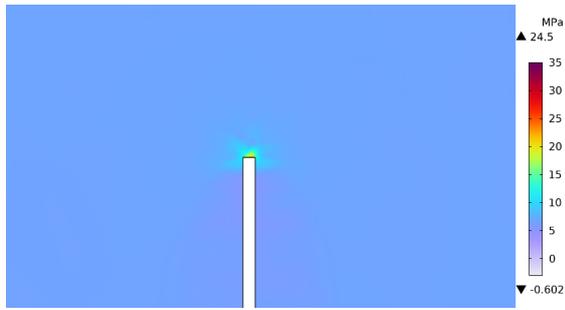

(e) Experimental configuration5：A$_2$B$_1$C$_2$

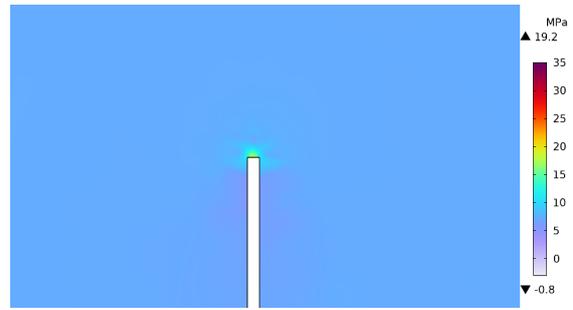

(f) Experimental configuration6：A$_2$B$_2$C$_1$

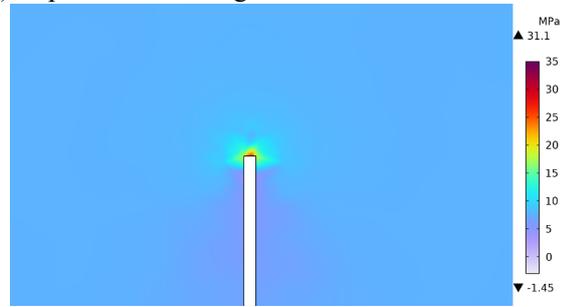

(g) Experimental configuration7：A$_2$B$_3$C$_4$

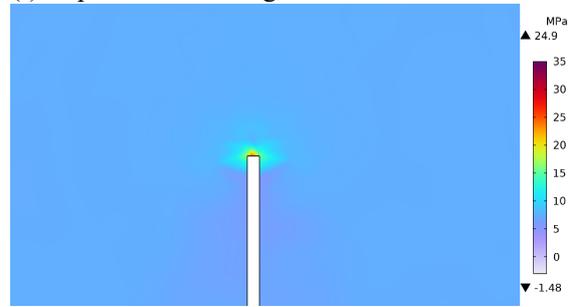

(h) Experimental configuration8：A$_2$B$_4$C$_3$

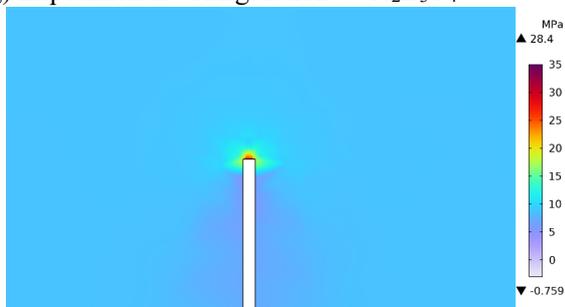

(i) Experimental configuration9：A$_3$B$_1$C$_3$

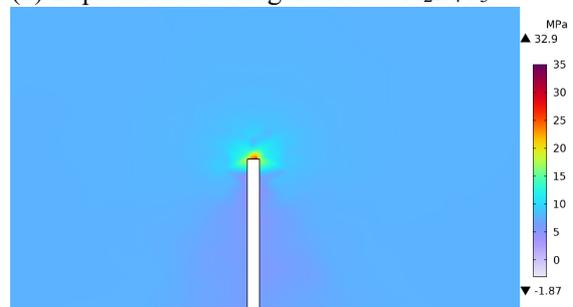

(j) Experimental configuration10：A$_3$B$_2$C$_4$

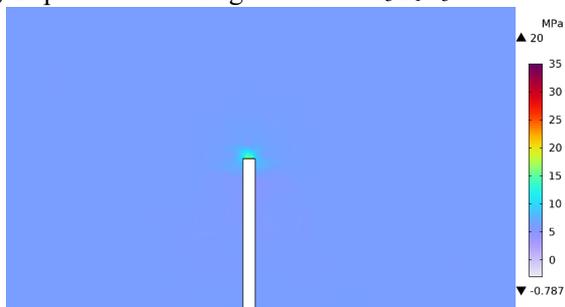

(k) Experimental configuration11：A$_3$B$_3$C$_1$

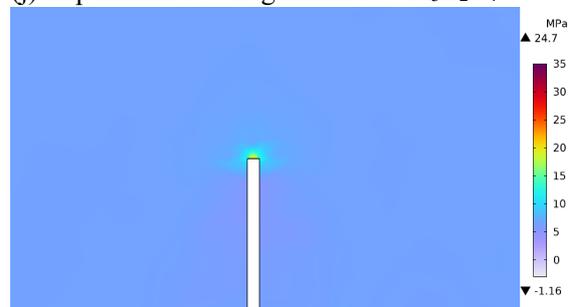

(l) Experimental configuration12：A$_3$B$_4$C$_2$

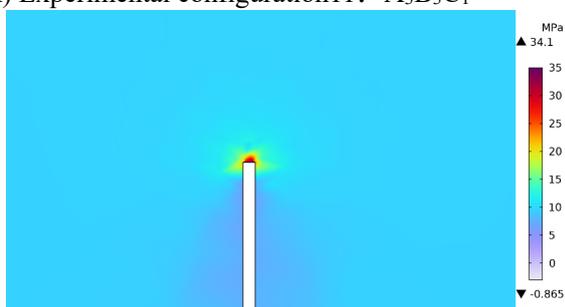

(m) Experimental configuration13：A$_4$B$_1$C$_4$

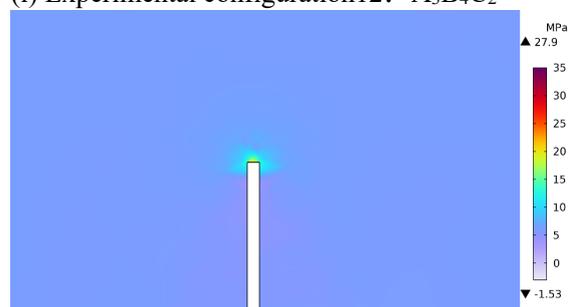

(n) Experimental configuration14：A$_4$B$_2$C$_3$

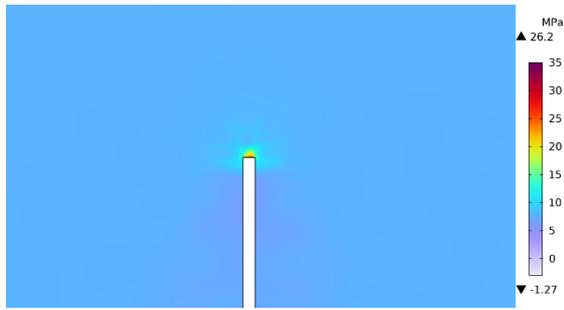
(o) Experimental configuration15: $A_4B_3C_2$

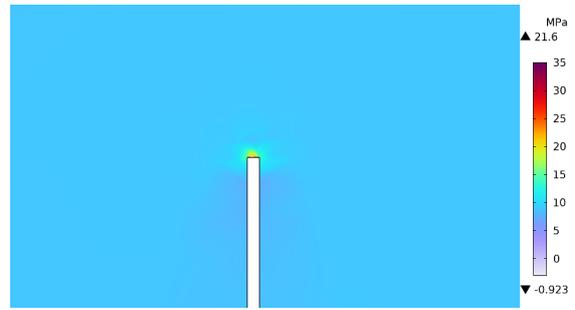
(p) Experimental configuration16: $A_4B_4C_1$

Fig 7 Maximum principal stress (σ1) distribution at the upper crack mouth

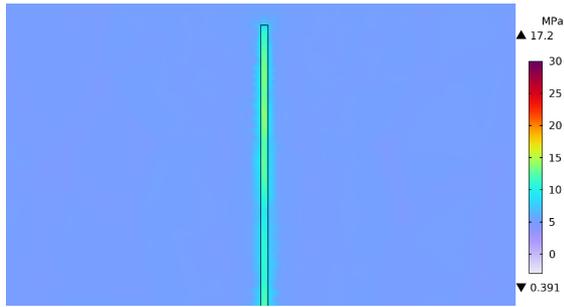
(a) Experimental configuration1: $A_1B_1C_1$

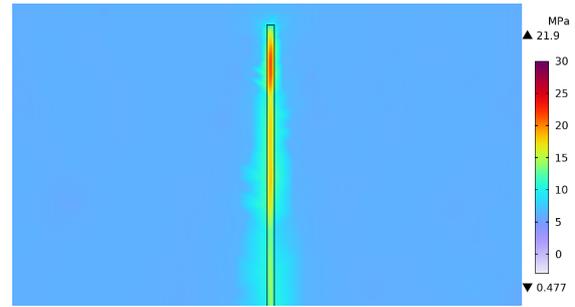
(b) Experimental configuration2: $A_1B_2C_2$

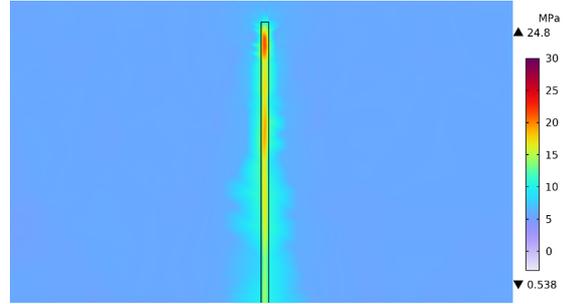
(c) Experimental configuration3: $A_1B_3C_3$

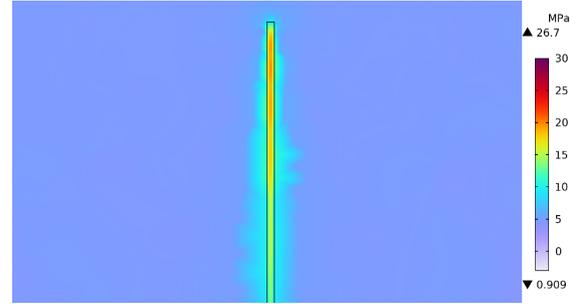
(d) Experimental configuration4: $A_1B_4C_4$

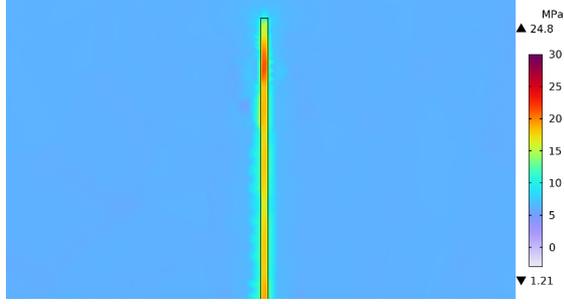
(e) Experimental configuration5: $A_2B_1C_2$

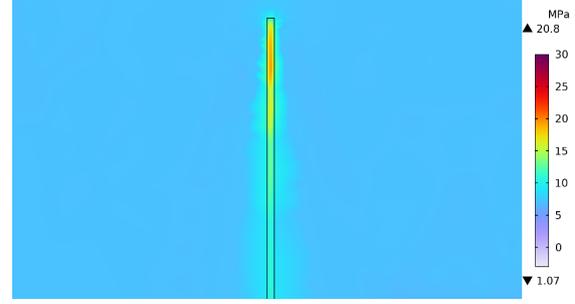
(f) Experimental configuration6: $A_2B_2C_1$

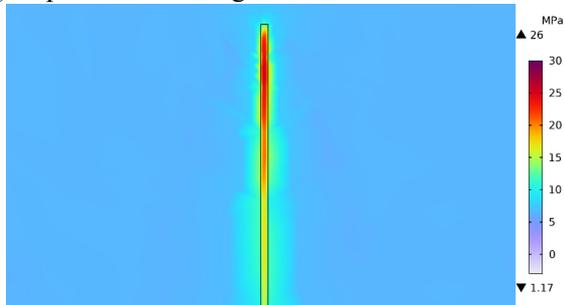
(g) Experimental configuration7: $A_2B_3C_4$

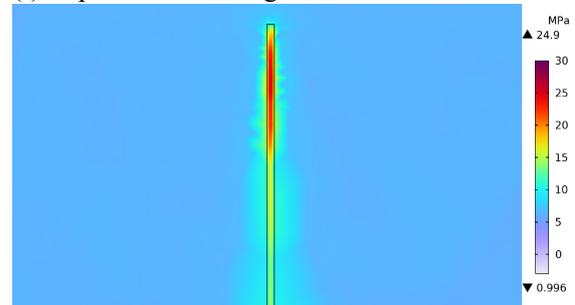
(h) Experimental configuration8: $A_2B_4C_3$

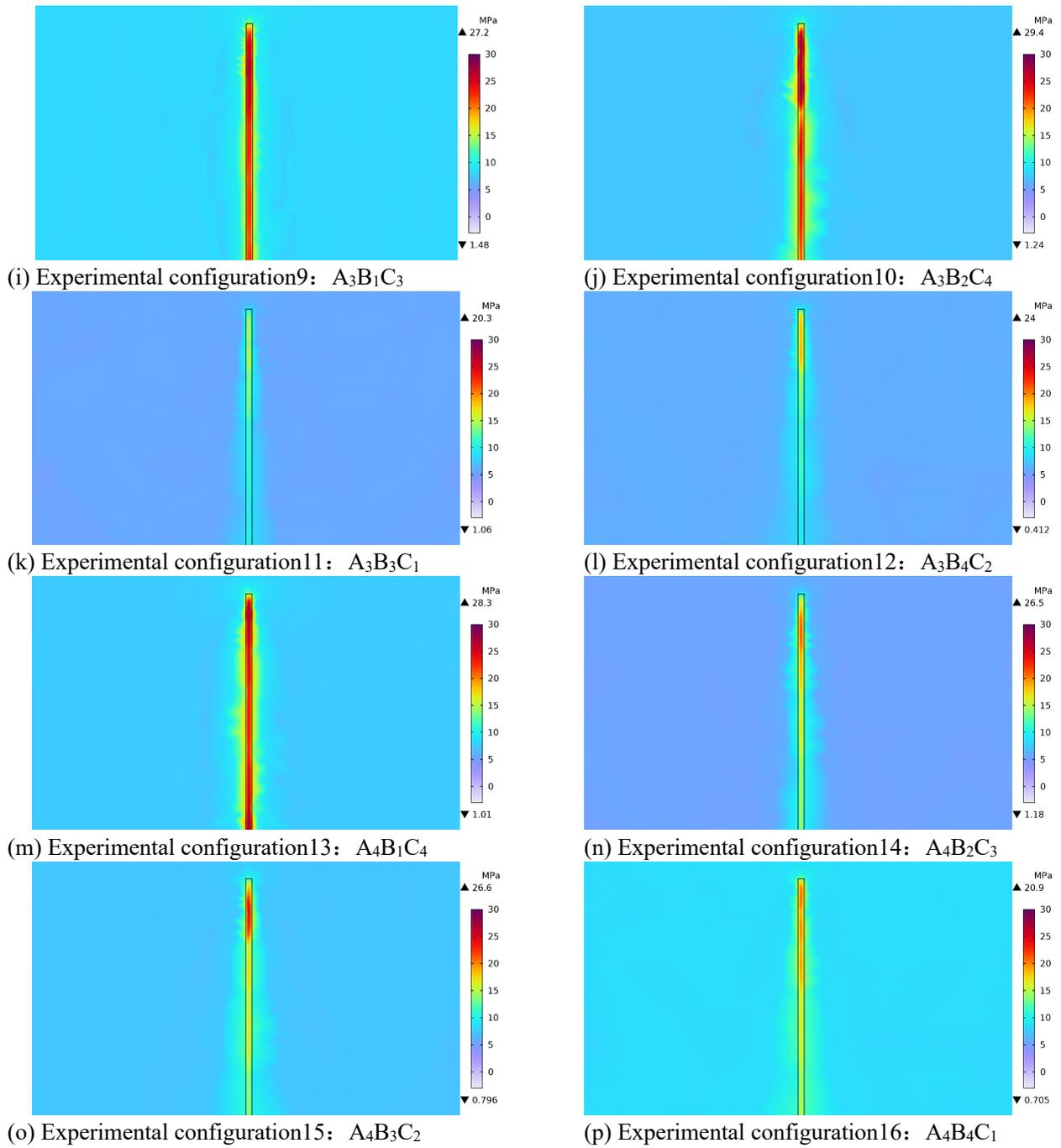

(i) Experimental configuration9: $A_3B_1C_3$

(j) Experimental configuration10: $A_3B_2C_4$

(k) Experimental configuration11: $A_3B_3C_1$

(l) Experimental configuration12: $A_3B_4C_2$

(m) Experimental configuration13: $A_4B_1C_4$

(n) Experimental configuration14: $A_4B_2C_3$

(o) Experimental configuration15: $A_4B_3C_2$

(p) Experimental configuration16: $A_4B_4C_1$

Fig 8 Maximum principal stress (σ1) distribution at the crack bottom

As shown in the figures, the orthogonal simulation test reveals that the maximum first principal stress ranges from 19.2 MPa to 34.1 MPa at the crack surface end and from 17.2 MPa to 29.4 MPa at the crack bottom. These stress values are derived from steady-state analysis using a linear elastic constitutive model. The results show that when there are existing cracks in the concrete structure, under the combined action of seasonal temperature changes and constraints, obvious stress concentration phenomena will occur at the crack tip and its bottom. During the stage of seasonal temperature difference caused by "forming at high temperature in summer and serving at low temperature in winter", the stress value in this concentrated area is significantly higher than the stress level at the same position in the case of no cracks.

*3.3. Orthogonal experimental factor sensitivity analysis*

Select the maximum tensile stresses (i.e., the maximum values of the first principal stress) at the surface end and the bottom of the crack in the simulation results of 16 combinations for analysis, as shown in the Fig 9.

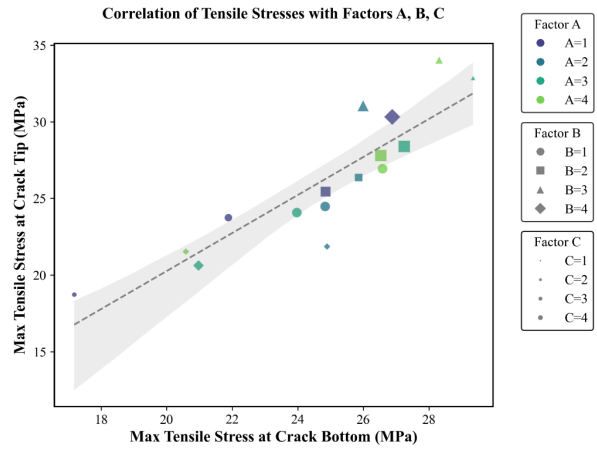

Fig 9. Results of the orthogonal experiment considering the average Forming temperature

The range analysis method[21] is the most commonly used and intuitive method for judging the significance of factors in orthogonal tests. The larger the range of a factor is, the greater the change in the target parameter is when the factor changes within the test range. Its calculation formula is as follows[22]:

$$k_i = (1/s) \times \sum_{k=1}^{s} y_k \quad (4)$$

Where:
$k_i$ ---the arithmetic means when the factor takes level;
$s$ — the number of trials at level i of this factor;
$R$ — the range of this factor.

(1) End-part Factor Sensitivity Analysis

Conduct a sensitivity analysis of the crack - end stress for different factors. The mean values and ranges of different levels of each factor are shown in Table 4.

Table 4
Range analysis of the maximum tensile stress at the end

| Level | Factor | | |
|---|---|---|---|
| | A | B | C |
| Mean value of the level k1 | 24.57 | 26.41 | 19.88 |
| Mean value of the level k2 | 25.27 | 25.86 | 24.7 |
| Mean value of the level k3 | 26.49 | 25.75 | 27.1 |
| Mean value of the level k4 | 27.42 | 25.73 | 32.08 |
| Range | 2.85 | 0.68 | 12.2 |
| Rank | 2 | 3 | 1 |

It can be seen from the range (R) and ranking results in the Table 5 that, regarding the magnitude of the maximum tensile stress at the crack end, the sensitivity order of each factor is (C > A > B). That is, from the perspective of the strength of the influence, the crack depth has the greatest influence, followed by the initial temperature, and then the crack length. Next, the effects of different levels of each factor will be further studied.

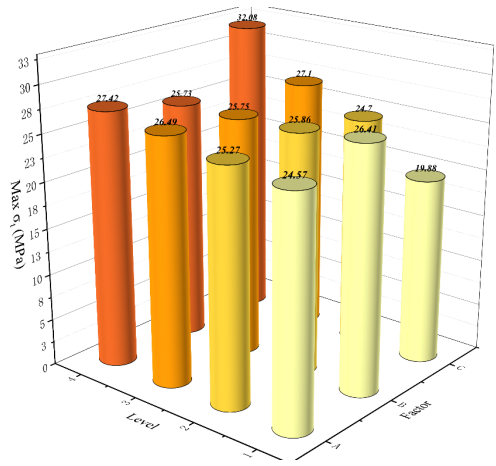

Fig 10. Mean value of each factor level for the maximum tensile stress at the crack end

It can be known from the analysis of the Fig 10 that the value corresponding to level 4 of factor A is the largest, the value corresponding to level 1 of factor B is the largest, and the value corresponding to level 4 of factor C is the largest. Therefore, the unfavorable combination is $(A_4 B_1 C_4)$, that is, this combination of factors and levels has the greatest influence on the maximum tensile stress at the crack end. However, the gap between each level of factor B is very small. Therefore, when the crack length is determined, the higher the temperature when pouring into the mold and the greater the crack depth, the larger the value of the maximum tensile stress at the crack end, and the crack end of the concrete is more likely to propagate.

(2)Bottom-part Factor Sensitivity Analysis

Similarly, the sensitivity analysis of the stress at the bottom of the crack is carried out, as shown in Table 5

Table 5
Range analysis of the maximum tensile stress at the bottom

| Level | Factor | | |
| --- | --- | --- | --- |
| | A | B | C |
| Mean value of the level k1 | 22.65 | 24.39 | 19.82 |
| Mean value of the level k2 | 24.13 | 24.65 | 24.31 |
| Mean value of the level k3 | 25.23 | 24.44 | 25.87 |
| Mean value of the level k4 | 25.59 | 24.11 | 27.59 |
| Range | 2.95 | 0.54 | 7.76 |
| Rank | 2 | 3 | 1 |

It can be seen from the range (R) and ranking results in the table5 that, regarding the magnitude of the maximum tensile stress at the bottom of the crack, the sensitivity order of each factor is (C > A > B). That is, from the perspective of the strength of the influence, the crack depth has the greatest influence, followed by the temperature when pouring into the mold, and then the crack length. This is consistent with the ranking of the influencing factors of the maximum tensile stress at the crack end. Next, the effects of different levels of each factor will also be further studied.

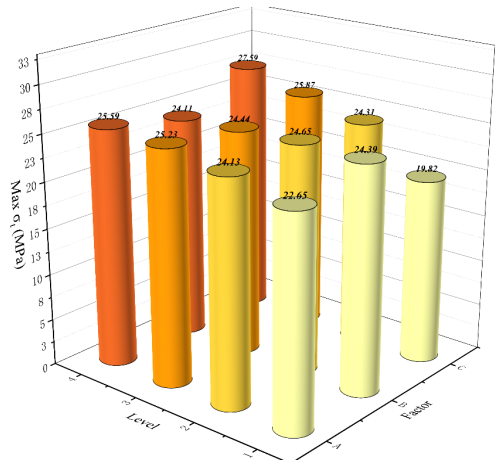

Fig 11. Mean value of each factor level for the maximum tensile stress at the bottom of the crack

It can be known from the analysis of the Fig 11 that the value corresponding to level 4 of factor A is the largest, the value corresponding to level 2 of factor B is the largest, and the value corresponding to level 4 of factor C is the largest. Therefore, the most unfavorable combination is $(A_4B_2C_4)$, that is, this combination of factors and levels has the greatest influence on the maximum tensile stress at the bottom of the crack. However, similarly, the gap between each level of factor B is very small. Therefore, when the crack length is determined, the higher the temperature when pouring into the mold and the greater the crack depth, the larger the value of the maximum tensile stress at the bottom of the crack, and the bottom of the concrete crack is more likely to propagate.

In conclusion, based on the analysis of the influence of the magnitude of the tensile stress at the end and bottom of the existing cracks in the concrete under the influence of the external low temperature, the influence of the two factors, namely the depth of the existing cracks and the temperature of the concrete when it is poured into the mold, is higher than that of the crack length. The existing crack length has almost no influence, and the influence of the crack depth is the most obvious, which indicates that the greater the crack depth, the easier the crack is to propagate under the influence of the external temperature. Moreover, in order to avoid the Forming of cracks on the surface of the early-stage concrete, the temperature of the concrete when it is poured into the mold should not be too high.

## 4. Discussion

### 4.1. Analysis of Internal Defect Results

In order to more clearly compare the stress differences on the surface of the holes at different temperatures, Fig 12 to Fig 15 show the probability density distribution diagrams of the stress on the surface of the holes with an interval of 0.05 MPa within the range of -1 MPa to 7 MPa for the first principal stress under different longitudinal dimensions.

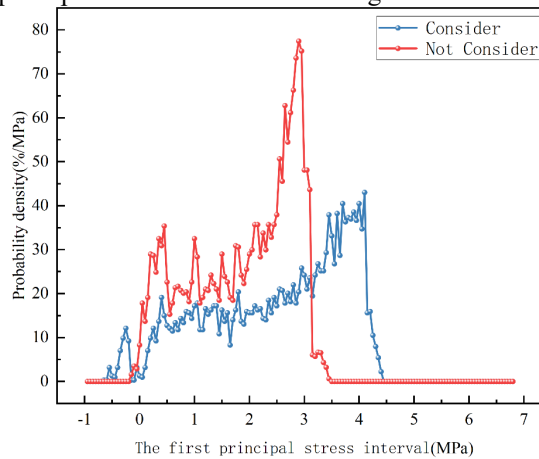

Fig 12. Ambient temperature: 10°C: Probability density distribution diagram of the first principal stress;

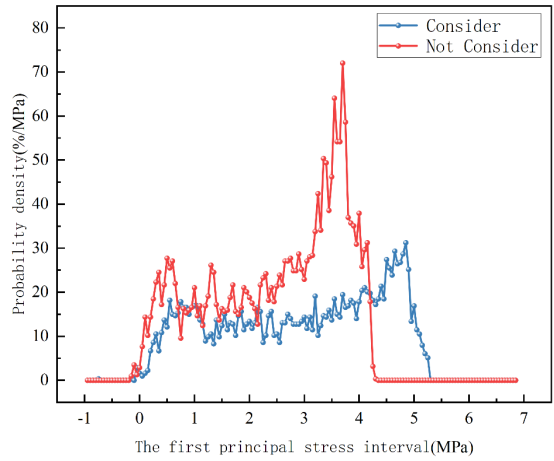

Fig 13. Ambient temperature: 5°C: Probability density distribution diagram of the first principal stress;

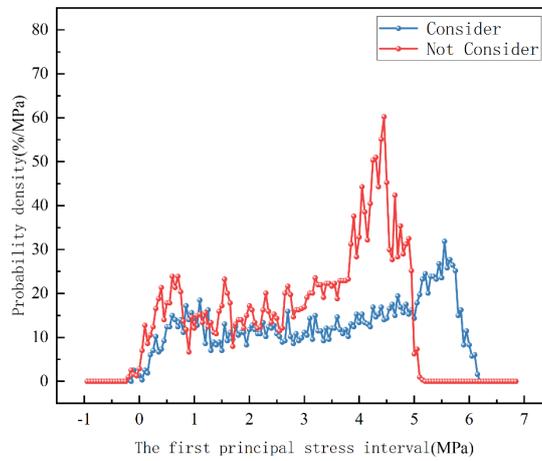

Fig 14. Ambient temperature: 0°C: Probability density distribution diagram of the first principal stress;

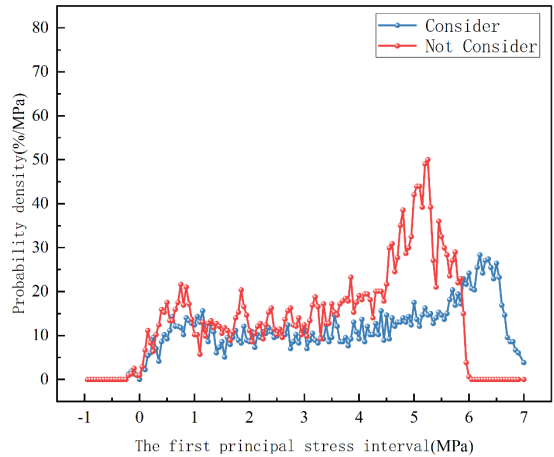

Fig 15. Ambient temperature: -5°C: Probability density distribution diagram of the first principal stress;

Through comparative analysis, we find that, at each ambient temperature (10, 5, 0, −5 °C), adopting the average forming temperature (AFT) yields a broader $\sigma_1$ distribution on void surfaces with a right-shifted peak relative to the non-AFT case, indicating consistently higher stress levels with AFT. As ambient temperature decreases, the peak in both cases shifts further to the right, i.e., lower temperature increases $\sigma_1$.

It can be seen from the data in the Fig 16 that regardless of whether the average forming temperature is considered or not, as the ambient temperature increases, the maximum and minimum values of the tensile stress show a downward trend. In other words, the lower the ambient temperature, the greater the stress value generated on the surface of the holes, which is more unfavorable for the concrete structure with internal defects. The difference between the two cases has remained constant, which indicates that the difference in the maximum stress values between the two cases does not change with the ambient temperature.

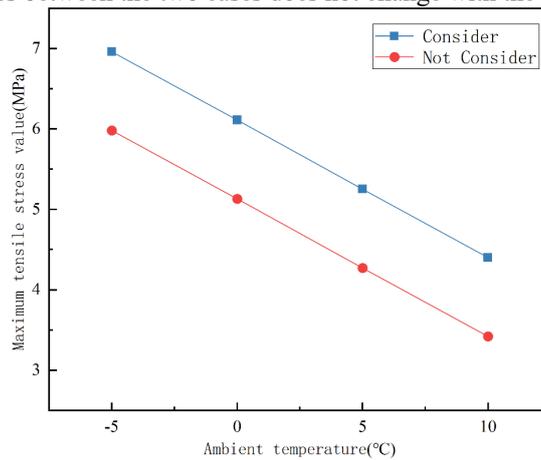

Fig 16. The maximum and minimum values of the tensile stress on the surface of the air voids under different ambient temperatures

### 4.2. Analysis of Surface Defect Results

The simulation results of the previous orthogonal test scheme were calculated based on the consideration of the AFT. In order to comprehensively evaluate the influence of considering the AFT on the seasonal temperature stress, comparative simulation without considering the AFT were further carried out in the same 16 orthogonal test combinations. In this case, the temperature of concrete placement is taken as the reference temperature. Similarly, the maximum tensile stress values at the surface port and bottom of the crack are extracted. The results are shown in the following Fig 17:

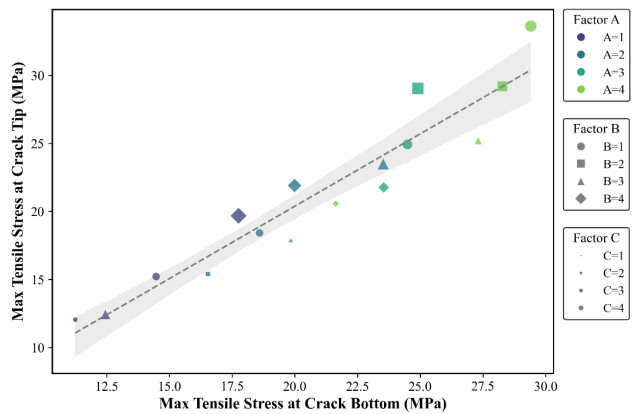

Fig 17. Results of the orthogonal experiment without considering the average Forming temperature

In order to more clearly compare the differences in stress distribution between the two cases of "considering the AFT" and "not considering the AFT", we subtract the simulation results of the two cases, and then analyze the differences under different working conditions. As shown in Fig 18:

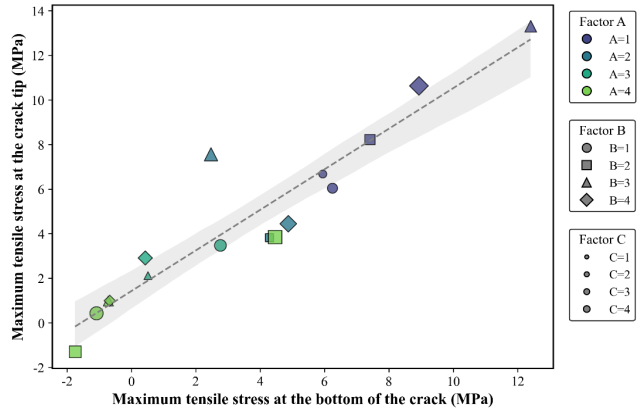

Fig 18. Stress differences between analyses with and without adopting the average forming temperature (AFT)

Note: The difference value represents the stress value when the AFT is considered minus the stress value when the AFT is not considered.

1)Analysis of the Sensitivity of Factors Affecting the Stress Difference at the End

The range analysis method is used to conduct a sensitivity analysis of the stress difference at the crack end for different factors. The data is grouped according to the different levels of the factors, and the average value of each factor at each level and the range of the mean values are calculated.

Table 6
Range analysis of the difference values at the crack end

| Level | Factor | | |
|---|---|---|---|
| | A | B | C |
| Mean value of the level k1 | 9.72 | 4.16 | 3.4 |
| Mean value of the level k2 | 5.47 | 3.65 | 4.54 |
| Mean value of the level k3 | 3.09 | 6 | 5.99 |
| Mean value of the level k4 | 0.28 | 4.75 | 5.62 |
| Range | 9.44 | 2.35 | 2.59 |
| Rank | 2 | 3 | 3 |

It can be seen from the range (R) and ranking results in the Table 6 that the range of the temperature of concrete placement is the largest, indicating that it has the most significant influence on the stress difference at the crack end under the two conditions. It also shows that the temperature of concrete placement is a key factor affecting the temperature field of concrete during the forming period. As the temperature of concrete placement increases, the mean value at each level of placement temperature gradually decreases. This indicates that at a higher temperature of concrete placement, the influence of the ambient temperature during the forming period on the concrete temperature field is weakened, causing the AFT to gradually approach the temperature of concrete placement. As a result, after considering the AFT, the difference in the end stress value from that without considering it is reduced. In contrast, the range values of the crack length and crack depth are small and close to each other, indicating that their influence on the temperature field around the end of the concrete crack is relatively limited. Therefore, compared with the temperature of concrete placement, the influence of crack parameters on the stress difference under the two conditions is weak.

(2)Analysis of the Sensitivity of Factors Affecting the Stress Difference at the Bottom

Similarly, the range analysis method is used to conduct a sensitivity analysis of the stress difference at the bottom of the crack for different factors.

Table 7
Range analysis of the difference values at the bottom of the crack

| Level | Factor | | |
|---|---|---|---|
| | A | B | C |
| Mean value of the level k1 | 8.67 | 3.46 | 2.52 |
| Mean value of the level k2 | 4.47 | 3.6 | 3.34 |
| Mean value of the level k3 | 2.04 | 3.66 | 4.57 |
| Mean value of the level k4 | -1.06 | 3.39 | 3.36 |
| Range | 9.73 | 0.27 | 2.05 |
| Rank | 1 | 3 | 3 |

It can be seen from the range (R) and ranking results in the Table 7 that, again, the range of the temperature of concrete placement is the largest, which has the most significant influence on the stress difference at the bottom under the two conditions. Next is the crack depth, while the influence of the crack length is negligible. This result is basically consistent with the conclusion of the range analysis of the difference at the crack end. In addition, as the temperature of concrete placement increases, the horizontal average value also shows a gradually decreasing trend, and negative values appear. This indicates that the influence of the ambient temperature leads to a situation where the AFT is lower than the temperature of concrete placement. In contrast, the range value of the crack depth is small, indicating that this factor has a relatively limited influence on the temperature field at the bottom of the concrete crack; the range value of the crack length is close to 0, indicating that this factor has almost no influence on the temperature field around the bottom of the concrete crack. Therefore, compared with the temperature of concrete placement, the influence of crack parameters on the stress difference under the two conditions is still weak.

In conclusion, when the AFT is considered, the stress value around the crack is generally higher than that when it is not considered. The temperature of concrete placement has a significant influence on the stress difference between the two situations, while the changes in crack length and depth have a relatively small impact on the stress difference. Although the temperature of concrete placement has a significant influence on the stress difference between the two cases, as the temperature of concrete placement increases, this difference gradually decreases, and it may even lead to the situation where the stress value when the AFT is considered is lower than that when it is not considered.

## 5. Conclusion

This study establishes the significance of incorporating the average forming temperature (AFT) as the strain-free reference in thermal–stress assessment of defective mass concrete. For internal voids, considering AFT consistently elevates the first principal tensile stress on void surfaces relative to analyses that neglect it; under identical cold environments the peak tensile stress rises from 3.42 to 4.40 MPa at 10 °C and from 5.98 to 6.96 MPa at −5 °C. Probability–density comparisons show a rightward shift and broadening of the stress distribution when AFT is included, while the peak–to–peak between the two scenarios remains constant. These results indicate higher cracking propensity at low temperatures and confirm that AFT matters even when the ambient-temperature effect dominates.

For surface defects, orthogonal analyses (L16(4^3), three factors × four levels) demonstrate that crack depth is the primary driver of tensile stress at both the crack tip and the crack bottom, followed by concrete placement temperature, with crack length least influential (depth > placement temperature > length). Greater depth markedly amplifies peak tensile stress, accelerating crack-growth risk.

Comparative simulations further reveal that the stress increment attributable to AFT is governed mainly by the placement (molding) temperature, exceeding the influence of defect length and depth on the difference between "with-AFT" and "without-AFT" cases (molding temperature > depth > length). As placement temperature increases, the two predictions converge and the AFT-induced increment diminishes, and may even reverse in isolated combinations—underscoring placement temperature as a critical input for AFT-based stress evaluation.

Practically, the findings recommend:
(i) explicitly adopting AFT as the strain-free field in winter checks;
(ii) constraining placement temperature to limit AFT-driven tensile amplification;
(iii) prioritizing identification and mitigation of deep surface cracks. These measures provide an actionable basis for temperature-control and defect-management strategies in mass-concrete construction and operation.

**Acknowledgments**